\documentclass[conference,twocolumn, twoside]{IEEEtran}
\usepackage{amssymb,epsfig,graphics, graphicx, url,times,mathrsfs,algorithm,algorithmic,amsmath,array,latexsym,fancyhdr,xspace,wrapfig, xcolor,multirow,amsthm,cite,helvet,sidecap,bm,comment,upref}
\usepackage[normalem]{ulem}
\usepackage{balance}
\usepackage[inline]{enumitem}
\usepackage{arydshln}
\usepackage{multirow}
\usepackage[makeroom]{cancel}
\usepackage{varwidth}
\usepackage{pgfplots}
\usetikzlibrary{spy}
\usepackage{ctable}
\usepackage{tabularx}

\usepackage{tikz}
\usepackage{nicefrac}
\usetikzlibrary{shapes.geometric}
\usetikzlibrary{shapes,snakes,patterns}
\usetikzlibrary{arrows,positioning,automata,calc}
\usetikzlibrary{plotmarks}

\definecolor{green1}{rgb}{0,0.5,0}
\definecolor{magenta}{rgb}{1.0, 0.11, 0.81}
\definecolor{mulberry}{rgb}{0.77, 0.29, 0.55}
\definecolor{xgray}{rgb}{0.9, 0.9, 0.9}

\def \bes{\begin{equation*}}
\def \ees{\end{equation*}}
\def \bas{\begin{align*}}
\def \eas{\end{align*}}
\def \be{\begin{equation}}
\def \ee{\end{equation}}
\def \bbm{\begin{bmatrix}}
\def \ebm{\end{bmatrix}}

\newcommand{\cD}{{\cal D}}
\newcommand{\cE}{{\cal E}}
\newcommand{\cF}{{\cal F}}
\newcommand{\cG}{{\cal G}}

\newcommand{\cI}{{\cal I}}

\newcommand{\cK}{{\cal K}}

\newcommand{\cS}{{\cal S}}
\newcommand{\cT}{{\cal T}}

\newcommand{\bfc}{{\boldsymbol c}}

\newcommand{\bfe}{{\boldsymbol e}}

\newcommand{\bfm}{{\boldsymbol m}}

\newcommand{\bfs}{{\boldsymbol s}}

\newcommand{\bfx}{{\boldsymbol x}}

\newcommand{\bfA}{{\boldsymbol A}}

\newcommand{\bfC}{{\boldsymbol C}}
\newcommand{\bfD}{{\boldsymbol D}}
\newcommand{\bfE}{{\boldsymbol E}}

\newcommand{\bfG}{{\boldsymbol G}}
\newcommand{\bfH}{{\boldsymbol H}}

\newcommand{\bfR}{{\boldsymbol R}}
\newcommand{\bfS}{{\boldsymbol S}}
\newcommand{\bfT}{{\boldsymbol T}}

\newcommand{\bfX}{{\boldsymbol X}}

\newcommand{\bfTheta}{{\boldsymbol \vartheta}}

\newcommand{\code}{\mathcal{C}}

\usetikzlibrary{decorations.pathreplacing,calc}
\tikzset{brace/.style={decorate, decoration={brace}},
 brace mirrored/.style={decorate, decoration={brace,mirror}},
}

\newcounter{brace}
\definecolor{igreen}{rgb}{0.07, 0.53, 0.03}
\definecolor{battleshipgrey}{rgb}{0.52, 0.52, 0.51}

\newcommand{\transpose}{\mathsf{T}}

\newcommand{\bse}{\boldsymbol{e}}
\newcommand{\bH}{\boldsymbol{H}}
\newcommand{\bG}{\boldsymbol{G}}
\newcommand{\rE}{\text{E}}

\DeclareMathOperator{\supp}{supp}
\DeclareMathOperator{\wt}{wt}

\newcommand\scalemath[2]{\scalebox{#1}{\mbox{\ensuremath{\displaystyle #2}}}}

\IEEEoverridecommandlockouts

\begin{document}

\title{Computational Code-Based Privacy in\\ Coded Federated Learning\vspace{-.3cm}}
\author{\IEEEauthorblockN{
Marvin Xhemrishi$^1$, 
Alexandre Graell i Amat$^{2,3}$, 
Eirik Rosnes$^3$, 
and Antonia Wachter-Zeh$^1$\\
$^1$Institute of Communications Engineering, Technical University of Munich, Munich, Germany\\
$^2$Chalmers University of Technology, Gothenburg, Sweden\\
$^3$Simula UiB, Bergen, Norway\\
Emails: \{marvin.xhemrishi, antonia.wachter-zeh\}{@tum.de}, {alexandre.graell@chalmers.se}, {eirikrosnes@simula.no}
}
\thanks{M.~Xhemrishi's work is funded by a DFG (German Research Foundation) project under Grant Agreement No. WA 3907/7-1. 
 The work by A. Graell i Amat was supported by the Swedish Research Council under grant 2020-03687.}
}
\IEEEaftertitletext{\vspace{-2\baselineskip}}

\maketitle
\begin{abstract}
We propose a privacy-preserving federated learning (FL) scheme that is resilient against straggling devices. An adaptive scenario is suggested where the slower devices share their data with the faster ones and do not participate in the learning process. The proposed scheme employs  code-based cryptography to ensure \emph{computational} privacy of the private data, i.e., no device with bounded computational power can obtain information about the other devices' data in feasible time.  For a scenario with 25 devices, the proposed scheme achieves a speed-up of 4.7 and 4 for 92 and 128 bits security, respectively, for an accuracy of 95\% on the MNIST dataset compared with conventional mini-batch FL.  
\end{abstract}

\section{Introduction}\label{sec:Intro}

Federated learning (FL) is a  form of  machine learning that trains a global model on decentralized data. The key concept is that devices do not share their data with the central server learning the global model, but only the local gradients, which are aggregated by the central server to  update the  model. By keeping raw data local, some level of privacy is preserved.

FL suffers from \emph{straggling devices}, i.e.,  
low-processing devices, which may induce a high training latency. A simple way to  circumvent this shortcoming is to ignore straggling devices~\cite{FL_averaging}, i.e., training the global model only on the data of the fastest devices. This approach, however, may lead to the so-called client drift problem if the data is not identically distributed across devices, which is typically the case~\cite{async1, async_2}. 

Providing resilience against stragglers has been addressed in the sister field of distributed computing, where computations are distributed over many workers, by means of erasure correcting codes~\cite{speeding_up_using_codes, gradient_coding, poly_codes, albin_paper, factored_lt,short_dot}. The main principle is to introduce redundant computations such that the computation can be completed from the tasks of a subset of the workers. The work in \cite{Intel_CFL} was the first to adopt techniques from coded distributed computing in FL by  encoding the local data and offloading the parity data to the central server. However, sharing parity data with the central server leaks information, hence the scheme in \cite{Intel_CFL} yields a lower level of privacy than conventional FL. A coded FL scheme that preserves the same level of privacy as FL was proposed in~\cite{kumar2021coding}. The scheme in~\cite{kumar2021coding} introduces redundancy by sharing data between devices, which enables the use of gradient codes to provide straggler mitigation. To provide privacy, the data is one-time padded  before sharing  it with other devices.

In this paper, similar to~\cite{kumar2021coding}, we propose a coded FL scheme  that mitigates the impact of stragglers on the training latency by replicating data across devices, while preserving the same level of privacy as conventional FL. In contrast to~\cite{kumar2021coding}, which provides information-theoretic privacy on the data shared between devices, the proposed scheme yields \emph{computational} privacy, i.e., no device with limited computational power can learn anything about the data of other devices. In particular, devices encrypt their data using  a code-based cryptosystem  \cite{McEliece_original, Niederreiter_original} whose security relies on the hardness of decoding random codes and is robust against the attack of a quantum computer.  We use an \emph{interleaved} McEliece cryptosystem~\cite{lukaslia} that  is carefully designed  such that it is homomorphic under addition and multiplication by a constant, i.e., a computation over the encrypted data is mirrored in the plaintext data. 
We propose an adaptive data sharing technique where a straggling device transfers its data to a nonstraggling device. A device is considered to be a straggler if  it replies later than $\Delta$ seconds from the fastest device.
For linear regression over the MNIST dataset, the proposed FL scheme achieves a speed-up of $4.7$ and $4$ for a security level of $92$ and $128$ bits, respectively, for an accuracy of $95\%$  compared with  conventional mini-batch FL.

\emph{Notation}: We denote row vectors by lowercase bold letters,  matrices by uppercase bold letters, and sets by calligraphic letters, e.g., $\bfx$, $\bfX$, and $\mathcal{X}$, respectively. 
We use $\supp(\bfx)$ to denote  the support of a vector $\bfx$, and denote by $\wt(\bfx)$ its weight, i.e., the number of nonzero entries. The field of the real numbers is denoted by $\mathbb{R}$, while a finite field of size $q$ is denoted by $\mathbb{F}_q$. The cardinality of a set $\mathcal{X}$ is denoted by $\lvert \mathcal{X}\rvert$, while the row span of a matrix $\bfX$ is denoted by $\langle \bfX \rangle$. The  Frobenius norm of a matrix $\bfX$ is denoted by $\lVert \bfX \rVert_\text{F}$. We denote by $d(\mathcal{C})$ the minimum Hamming distance of a code $\mathcal{C}$. For any positive integer $p$, the $\ell_p$-norm of a vector $\bfx = (x_1,\ldots,x_N) \in \mathbb{R}^N$  is denoted by   
$||\bfx||_p = \left(\sum_{i=1}^{N} \rvert x_i\lvert^p\right)^{1/p}$,
where $|x_i|$ is the absolute value of $x_i$.
For the special case $p=0$,  the $\ell_0$-norm of $\bfx$, $||\bfx||_0$, is equal to its weight, i.e., $\wt(\bfx)$. {The geometric distribution with probability of success $p_\text{s}$ is defined as geo$(1-p_\text{s})$.} Finally, we define  $[a]\triangleq \{1,2,\dots, a\}$ for a positive integer $a$.

\section{Preliminaries: The McEliece Cryptosystem}

The McEliece cryptosystem is a public-key code-based cryptosystem which relies on the hard problem of decoding a random code. Assume that a transmitter wants to transmit a message vector $\bfm$. Then, the legitimate receiver generates a key pair, where the public key is the generator matrix $\bfG$ of an error correcting code that can correct any $t$ errors with high probability. The private key is a parity-check matrix that allows for an efficient decoding algorithm, which will be explained in the sequel. The ciphertext is then computed as $$\bfc = \bfm\bfG + \bfe\,,$$
where $\bfe$ is an error vector generated at random with $\wt(\bfe) = t$. Extracting the plaintext message $\bfm$ from the ciphertext $\bfc$ without knowledge of the private key is a very hard problem. 
In this work, we consider a code-based cryptosystem based on moderate-density parity-check (MDPC) codes, which have recently been proposed for code-based cryptosystems~\cite{Tillich_MPDC}. MDPC codes are a class of codes-on-graphs closely related to the well-known low-density parity-check  codes, with the main difference that they are characterized by a slightly denser parity-check matrix~\cite{ Tillich_MPDC, QC-MDPC}.

The parity-check matrix, $\bH$, which allows the legitimate receiver to perform low-complexity decoding (e.g., via belief propagation) due to its medium density, conforms the private key. The receiver obtains a valid high-density generator matrix corresponding to $\bH$, denoted by $\bfG$, and it publishes it as its public key. It is computationally difficult to obtain a nondense parity-check matrix from $\bG$, hence an attacker cannot decode the ciphertext efficiently, and the system is computationally secure.

\section{System Model}\label{sec:sys_model}
We consider an FL scenario with $N$ devices and a central server. 
Device $i \in [N]$ has its own \emph{local data} $\cI_i = \left\{ (\boldsymbol{x}_j^{(i)}, \boldsymbol{y}_j^{(i)}) \vert j \in [n_i]\right\}$, consisting of $n_i$ \textit{features} and \textit{labels}. 
We denote by $m$ the number of data points across all devices, i.e., $m = \sum_{i=1}^N n_i$.   
The data can be represented in matrix form as 
\begin{equation*}
\boldsymbol{X}^{(i)} = \begin{pmatrix} \boldsymbol{x}_1^{(i)} \\ \vdots \\ \boldsymbol{x}_{
n_i}^{(i)} \end{pmatrix} \text{ and }\;  \boldsymbol{Y}^{(i)} = \begin{pmatrix} \boldsymbol{y}_1^{(i)} \\ \vdots \\ \boldsymbol{y}_{n_i}^{(i)} \end{pmatrix},
\end{equation*}
where $\boldsymbol{X}^{(i)}$ and $\boldsymbol{Y}^{(i)}$ are 
of dimensions $n_i\times d$  and  $n_i\times c$, respectively. 

\subsection{Federated Synchronous Gradient Descent}

We are interested in finding a suitable linear global model $ \boldsymbol{\Theta}$ of the form $$\boldsymbol{y} = \boldsymbol{x} \boldsymbol{\Theta}\,.$$
Finding this linear model can be achieved by solving the  minimization problem 
$$ \operatorname*{arg\,min}_{\boldsymbol{\Theta}} f(\boldsymbol{\Theta}) \triangleq \dfrac{1}{2m} \sum_{i=1}^{N}\sum_{j=1}^{n_i} \lVert \boldsymbol{x}^{(i)}_j\boldsymbol{\Theta} - \boldsymbol{y}^{(i)}_j\rVert_2^2 + \dfrac{\lambda}{2} \lVert\boldsymbol{\Theta}\rVert_{\text{F}}^2\,,$$
where $f(\boldsymbol{\Theta})$ is the global loss function and $\lambda$ the regularization parameter. Let the local loss function corresponding to the local data of device $i$ be  $$f_i(\boldsymbol{\Theta}) = \dfrac{1}{2n_i} \sum_{j=1}^{n_i} \lVert \boldsymbol{x}_j^{(i)}\boldsymbol{\Theta} - \boldsymbol{y}_j^{(i)}\rVert_2^2\,.$$Then, the global loss function can be rewritten as 
$$ f(\boldsymbol{\Theta}) = \sum_{i=1}^N \dfrac{n_i}{m} f_i(\boldsymbol{\Theta}) + \dfrac{\lambda}{2}\lVert\boldsymbol{\Theta}\rVert_{\text{F}}^2\,.$$
The devices  compute the gradient of the local loss function at each epoch $e$, 
\begin{equation}\label{eq:partial_grad}
\boldsymbol{\nabla}_i^{(e)} \triangleq n_i \nabla_{\boldsymbol{\Theta}}f_i(\boldsymbol{\Theta}^{(e)}) = {\boldsymbol{X}^{(i) }}^\transpose\boldsymbol{X}^{(i)}\boldsymbol{\Theta}^{(e)} - {\boldsymbol{X}^{(i) }}^\transpose\boldsymbol{Y}^{(i)}\,,
\end{equation}
where $\boldsymbol{\Theta}^{(e)}$ is the model estimate at epoch $e$, broadcasted by the server, and send them to the central server. The central server aggregates the received partial gradients to obtain the global gradient 
$$ \nabla_{\boldsymbol{\Theta}}f(\boldsymbol{\Theta}^{(e)}) = \boldsymbol{\nabla}^{(e)} + \lambda\boldsymbol{\Theta}^{(e)} = \sum_{i=1}^N \dfrac{1}{m} \boldsymbol{\nabla}_i^{(e)} + \lambda\boldsymbol{\Theta}^{(e)} $$
and  updates the model as 
\begin{equation}\label{eq:update_of_model}
\boldsymbol{\Theta}^{(e+1)} = \boldsymbol{\Theta}^{(e)} - \mu \nabla_{\boldsymbol{\Theta}}f(\boldsymbol{\Theta}^{(e)})\,,
\end{equation}
where $\mu$ is the learning rate.
Note that the second term in~\eqref{eq:partial_grad} is independent of the epoch $e$. Thus, we can write 
\begin{equation}\label{eq:other_epochs}
\boldsymbol{\nabla}_i^{(e)} = \boldsymbol{\nabla}_i^{(1)} + {\boldsymbol{X}^{(i) }}^\transpose\boldsymbol{X}^{(i)}\bfTheta^{(e)}\,,
\end{equation}
where $\bfTheta^{(e)} = \boldsymbol{\Theta}^{(e)} - \boldsymbol{\Theta}^{(1)}$. At the very first epoch, the server  broadcasts the initial model. For the successive epochs, the server broadcasts $\bfTheta^{(e)}$. For later use, we define 
\begin{equation}\label{eq:A_i}
    \bfA_i \triangleq {\bfX^{(i)}}^\transpose\bfX^{(i)}\,.
\end{equation}
$\bfA_i$ is a symmetric matrix and does not change throughout the epochs. Therefore, the computation performed by the devices to update the model is the  matrix-matrix multiplication $\boldsymbol{A}_i\bfTheta^{(e)}$. 

\subsection{{Computation and Communication Delay Model}}\label{subsec:random_setup_time}
{We model the computation times of the devices with a deterministic part and a random part. The deterministic part depends only on the computation power of the device, while the random component represents the time needed to start the computation (setup time). Particularly, let $\tau_i$ be the number of multiply and accumulate (MAC) operations that device $i$ can perform per second. The time required by device $i$  to perform $\rho_i$ MAC operations is then
$$T_i^\text{c} = \dfrac{\rho_i}{\tau_i} + \Lambda_i\,,$$
where $\Lambda_i$ is the random setup time, {which we model as an exponential random variable with rate $\zeta_i$  \cite{Simeone_letter}\textemdash $\zeta_i$ depends  on the computation power of the device.}

The devices communicate with the central server using a public channel. The communication link might fail and retransmissions are permitted. Let $P_i^\text{u}$ and $P_i^\text{d}$ be the number of trials until  device $i$ successfully transmits (uploads) or receives (downloads) a packet, respectively. We assume that communication  to/from device $i$ fails with probability $p_i$. Then,  $P_i^\text{u}$ and  $P_i^\text{d}$ follow the geometric distribution $\text{geo}(1-p_i)$, and the time needed to successfully upload and download $b$ bits is 
$$T_i^\text{u} = \dfrac{P_i^\text{u}}{\gamma^\text{u}}b \text{ and }T_i^\text{d} = \dfrac{P_i^\text{d}}{\gamma^\text{d}}b\,,$$respectively, 
where $\gamma^\text{u}$ and $\gamma^\text{d}$ denote the bit rate of the upload and download communication link, respectively. Moreover, the devices communicate over device-to-device (D2D) links of bit rate $\gamma^{\text{D2D}}$. The time needed between two devices to transmit $b$ bits using the D2D link is $$T^{\text{D2D}} = \dfrac{P^{\text{D2D}}}{\gamma^{\text{D2D}}}b\,,$$where $P^{\text{D2D}}$ follows the geometric distribution geo$(1-p^{\text{D2D}})$ with $p^{\text{D2D}}$ being the probability of a transmission failure in the D2D link.}

\section{Computationally-Private Federated Learning} \label{sec:CPFL}
In this section, we present our proposed  straggler-resilient FL scheme. 
To achieve straggler resiliency, we introduce redundancy by allowing 
devices to share their data with other devices. To yield privacy, the key idea is to encrypt the data using the McEliece cryptosystem prior to sharing it. The devices then  compute the local gradients on the encrypted data and the central server aggregates the received encrypted gradients to update the model. We describe in the following the three phases of the proposed scheme: encryption, data sharing, and computation. 

\subsection{Encryption}

We consider encryption via an MDPC code of  dimension~$k$ and length $n$. Let  $\boldsymbol{G} \in \mathbb{R}^{k\times n}$ be a generator matrix of the MDPC code, {which constitutes the public key}. For reasons that will be clarified later, our scheme uses an adaptive data sharing strategy in which   only a subset of the devices $\cD \subset [N]$, $\lvert \cD \rvert = D$,  encrypt their data and share it with other devices.
Device $u \in \cD$ encrypts its data matrix $\bfA_u$ before sharing it as 
\begin{equation}\label{eq:encryption_of_matrix}
    \boldsymbol{C}_u = \begin{pmatrix} \boldsymbol{c}_1^{(u)} \\ \vdots \\ \boldsymbol{c}_d^{(u)} \end{pmatrix} = \boldsymbol{A}_u\boldsymbol{G} + \boldsymbol{E}_u = \begin{pmatrix} \boldsymbol{a}_1^{(u)} \\ \vdots \\ \boldsymbol{a}_d^{(u)} \end{pmatrix}\boldsymbol{G} + \begin{pmatrix} \boldsymbol{e}_1^{(u)} \\ \vdots \\ \boldsymbol{e}_d^{(u)} \end{pmatrix},
\end{equation}
where $\boldsymbol{e}_j^{(u)} =(e_{j,1}^{(u)}, e_{j,2}^{(u)}, \dots, e_{j,n}^{(u)})$ and $e_{j,z}^{(u)} \neq 0  \text{ if } z \in \cE_u$ and $e_{j,z}^{(u)} = 0  \text{ if } z \notin \cE_u$, where $\cE_u$ is the error support designated for device $u$.
The encryption is done row-wise, i.e., each row of the resulting encrypted matrix is {a vector obtained as the sum of a codeword of the MDPC code and an error vector ($\bse_1^{(u)}$ to $\bse_d^{(u)}$). All rows share} the same error support $\cE_u$, which is decided and sent by the {central server} via a secure channel to device $u$ (this  channel can similarly be implemented by encrypting the support $\cE_u$ with the public key of device $u$).
The central server will use $\cE_u$, $u=1,\dots,D$, to employ erasure decoding in the decryption.

As explained later, the devices can add ciphertexts of different devices. This sum of ciphertexts  consists of codewords of the MDPC code that are corrupted at the union of the corresponding error supports. The server has then to correct this number of erasures. To guarantee that the desired data matrix can be reconstructed by the central server, we limit the size of $\cE \triangleq \cup_{i=1}^D \cE_i \subseteq [n]$ to $\lvert \cE \rvert = t_{\text{tot}}$. 
Simultaneously, we have to guarantee that device $i$ does not know too much about $\cE_j$ for all $j\in \{[D]\setminus i \}$. Otherwise, it could reduce the number of errors in $\boldsymbol{C}_j$ and possibly decrypt $\boldsymbol{A}_j$, violating the privacy constraints.
We therefore choose the $\cE_i$'s such that their pairwise intersections are small. In particular, for choosing $\cE$ and $\cE_i$, $i=1,\dots,N$, we proceed as follows:
i) Pick $\cE$ of size $t_{\text{tot}}$ uniformly at random from $[n]$;
ii) pick $D$ subsets $\cE_i \subseteq \cE$ of size $t \leq t_{\text{tot}}$ such that $\lvert\cE_i \cap \cE_j \rvert$ is ``small'' for all $i,j$. 

To guarantee ``small'' intersection, we note that the problem of finding {$M$} subsets $\cE_i$ of size $t$ from a set $\cE$ of size $t_{\text{tot}}$ with small intersection is equivalent to finding a binary constant-weight code of cardinality $M$, length $t_{\text{tot}}$, and weight $t$, where the $i$-th codeword, denoted by $\boldsymbol{w}_i$,  is $1$ at position $z$, $z \in [t_{\text{tot}}]$, if $z \in \cE_i$ and $0$ otherwise. Then, $|\cE_i \cap \cE_j|$ is equal to $t- \frac{d(\boldsymbol{w}_i,\boldsymbol{w}_j)}{2}$, where $d(\cdot, \cdot)$ denotes the Hamming distance between two words. Thus, a small intersection between the sets can be achieved by maximizing the minimum distance of the constant-weight code. For example, we can use the constant-weight codes from~\cite[Ex.~3]{CWC_Vardy}, see also Section~\ref{sec:num_res} for explicit choices of parameters. Let $M$ be the cardinality of the constant-weight code used in our scheme. Then, we are restricted to the case $D = M$. 

To illustrate that computation over encrypted data is possible, assume that at  epoch $e > 1$, the server sends $\bfTheta^{(e)}$ to a device that has the encrypted matrix $\bfC_u$. The device then performs the computation over encrypted data as 
\begin{align*}
\scalemath{0.93}{\widetilde{\bfC}_u^{(e)} = \bfTheta^{(e)^\transpose}\bfC_u= \bfTheta^{(e)^\transpose}\bfA_u\bfG + \bfTheta^{(e)^\transpose}\bfE_u  = (\bfTheta^{(e)^\transpose}\bfA_u)\bfG + \widetilde{\bfE}_u^{(e)}}\,,
\end{align*}
where $\widetilde{\bfE}_u^{(e)} = \bfTheta^{(e)^\transpose}\bfE_u$. The support $\widetilde{\mathcal{E}}_u$ of $\widetilde{\bfE}_u^{(e)}$ is the same as $\mathcal{E}_u$. As the number of erroneous positions does not increase and MDPC codes are linear, decoding $\widetilde{\bfC}_u^{(e)}$ will output $\bfTheta^{(e)^\transpose}\bfA_u$, whose transposition is what is needed to update the model. 

\subsection{Device-to-Device Communication}
We propose the following data sharing strategy. 
The server monitors the devices' response times throughout all the epochs of the learning process. 

If in a given epoch  device $j$ replies later than $\Delta$ seconds compared to the fastest device, the server sends to the device the error support $\cE_j$ and instructs it to encrypt the data (add the error matrix $\bfE_j$ to the encoded matrix) and send it to a faster device. Afterward, device $j$ is disregarded from the learning process and never contacted again. We denote the set containing the active workers at  epoch $e$ by $\mathcal{K}^{(e)}$ and its cardinality  by $K^{(e)}$. For the special case $e=1$, we have $\cK^{(1)} = [N]$ and $K^{(e)} = N$. 
Note that if $\Delta$ is small, the learning process may end up with a single device. 

Assume that at epoch $e$, device $i$ receives $\bfC_j$ from  device $j$, where $i\neq j$  and ${j \in \mathcal{D}}$. Then  device $i$ adds $\bfC_j$ with its encoded data to obtain $\bfD_i =(\bfA_i+\bfA_j)\bfG + \bfE_j$ and computes $\bfTheta^{(e)\transpose}\bfD_i$. Assume that at a later epoch $e' > e$, device $i$ is diagnosed as a straggler. 
If device $i$ shares $\bfD_i$ with another device, a critical privacy violation can occur, since device $j$ knows $\bfA_j\bfG +\bfE_j$. 
However, if device $i$ transmits $\bfD_i + \bfE_i$ and device $j$ eavesdrops, then it learn only  $t$ error-free positions. 
By picking $t$ carefully, we can ensure that finding out the other remaining erroneous positions is a hard problem. 
Ideally, $D = N$, which implies $M \geq N$. However, 
for some parameters, it is difficult to find constant-weight codes with $M\ge N$. Hence,  we consider  the case $M<N$.

{For every epoch $e$, we define by $L^{(e)}$ the number of available error supports, $L^{(1)} = M$. Without loss of generality, we assume $M$ is even. At an epoch $e$, the server identifies the set of straggling devices $\cS \subset \cK^{(e)}$. The server splits $\cS$ into two subsets, the subset of stragglers that were not recipient devices at previous epochs, $\cS^{\text{nr}}$, and the subset of straggling devices that received encrypted data from other devices in previous epochs, $\cS^{\text{r}}$ (for the first epoch, when stragglers are detected, it holds that $\cS^{\text{nr}} = \cS$). The server instructs the devices in $\cS^{\text{r}}$  to send their data without adding a new error matrix (since they have already two error supports and it is safe to do so) to faster devices denoted by $\cF^{\text{r}} \in \cK^{(e)}$, where $\lvert \cF^{\text{r}} \rvert = \lvert \cS^{\text{r}} \rvert $. The devices in $\cF^{\text{r}}$ are chosen as the slowest nonstraggling devices at epoch $e$ that were not recipient devices at previous epochs. If all the nonstraggling devices were recipient at previous epochs, the server picks as recipient devices the fastest ones at epoch $e$. Clearly if $ \lvert\cF^{\text{r}} \rvert \geq K^{(e)} - \lvert \cS \rvert $, then $\cF^{\text{r}} = \cK^{(e)} \setminus \cS$. For the case where $\lvert \cS^{\text{nr}} \rvert$ is even, the master selects a subset $\bar{\cS}^{\text{nr}} \subseteq {\cS}^{\text{nr}}$ of cardinality $\min\{\lvert \cS^{\text{nr}} \rvert, L^{(e)}\}$ that contains the slowest workers of $\cS^{\text{nr}}$ and sends each of them an unused error support (clearly, if $L^{(e)} =0$, it sends nothing and skips the data sharing for $\cS^{\text{nr}}$) and instructs them to add an error matrix to their encoded data. The server finds $\cF^{\text{nr}} \subseteq \cK^{(e)} $ of cardinality $\nicefrac{\lvert \bar{\cS}^{\text{nr}} \rvert}{2}$ recipient devices (same strategy as for $\cF^{\text{r}}$) and instructs them to send their data pair-wise, i.e., two straggling devices send to one recipient device. This way all  recipient devices in $\cF^{\text{nr}}$ will have at least $2$ error supports and will not require one in subsequent epochs. For the case where $\lvert \cS^{\text{nr}} \rvert$ is odd, the subset $\bar{\cS}^{\text{nr}}$ is picked of cardinality $\min\{\lvert \cS^{\text{nr}} \rvert +1, L^{(e)}\}$ and the same strategy follows. } 

\subsection{Computation} \label{subseq:computation}

To avoid multi-message communication between devices and the central server within an epoch, the devices encode their own data using the public key, i.e., $\bfG$. At epoch $e$,  device $v \in \cK^{(e)}$ will have $\Gamma_v^{(e)}  \in \{0,1,\dots, N-1\}$ data partitions other than its own data {and their identities are elements of the set $\cG_v^{(e)} $. The identities of the error matrices padded with the data are elements of the set $\cT_v^{(e)} \subseteq \cD$. }At epoch $e$, device $v \in \cK^{(e)}$ computes 
\begin{align*}
    \scalemath{0.87}{\bfT^{(e)}_v = \bfTheta^{(e)^\transpose} \left[ \left(\bfA_v + \sum_{i_v \in \cG_v^{(e)}} \bfA_{i_v} \right)\bfG + \sum_{i_u \in \cT_v^{(e)}} \bfE_{i_u} \right]
    = \bar{\bfA}_v^{(e)} + \bar{\bfE}_v^{(e)}}.
\end{align*}
By construction, $$\supp(\bar{\bfE}_v^{(e)}) = \bigcup\limits_{i_u \in \cT_v^{(e)}} \mathcal{E}_{i_u} \subseteq \cE\,. $$
After the computation has been performed, device $i$ sends $\bfT_i^{(e)}$ to the server and waits for the start of the new epoch. The server then waits for the $K^{(e)}$ active devices to reply back and aggregates the results to obtain $$\bar{\bfT}^{(e)} = \sum_{v \in \mathcal{K}^{(e)}} \bfT_v^{(e)}\,.$$It holds that $$\supp\bigl(\sum_{v \in \mathcal{K}^{(e)}} \bar{\bfE}_v^{(e)}\bigr) \subseteq \cE\,.$$
This allows the server to employ an erasure decoding strategy on $\bar{\bfT}^{(e)}$, e.g., peeling decoding~\cite{LT_codes} or inactivation decoding~\cite{Inactivation_Fran}, since by knowing the error positions the decoder can treat them as erasures.
{After successful decoding, the server retrieves the necessary computation 
$$\bfR^{(e)} = \bfTheta^{(e)^\transpose}  \sum_{i=1}^{N} \bfA_{i}\,,$$
which is an aggregation of all partial gradients. The server then sums  the first gradient with the aggregated gradient as  
$$\boldsymbol{\nabla}^{(e)} = \boldsymbol{\nabla}^{(1)} + \bfR^{(e)^\transpose}\,.$$ 
The server can now update the model in the $(e+1)$-th epoch as described in~\eqref{eq:update_of_model}. }

\section{Security Analysis}\label{sec:sec_analysis} 
Our encryption scheme can be seen as an interleaved scheme, since the matrices $\bfC_i$ contain erroneous codewords of an MPDC code as rows and the error support is the same for each row. {Thus, we consider attacks that apply to interleaved McEliece cryptosystems such as the one in~\cite{lukaslia}}. The number of interleaved codewords (number of rows) is called \emph{interleaving order}. Without loss of generality, we can focus on a single device, hence the device index $i$ is omitted in this section.

\subsection{Finding Low-Weight Codewords Attack} \label{subseq:Low_weight}
For the analysis of this low-weight codeword attack, see {\cite{lukaslia}}. Consider the  three {codebooks defined by the row spans}
\begin{align*}
\mathcal{C} \triangleq \langle \bfG \rangle \text{,  }\mathcal{C}' \triangleq \Bigg \langle \begin{bmatrix} \bfG \\ \bfC  \end{bmatrix} \Bigg \rangle  \text{, and } \mathcal{C}_{\rE} \triangleq \langle \bfE \rangle\,.
\end{align*}

By performing row operations, it is straightforward to see that $\mathcal{C}' = \mathcal{C} + \mathcal{C}_\rE$. Then, the minimum Hamming distance of $\mathcal{C}'$ is upper bounded by the minimum Hamming distance of the error code $\mathcal{C}_E$, i.e.,  $d(\mathcal{C'}) \leq d(\mathcal{C}_\rE) \leq t$. The code $\mathcal{C}'$ is composed of two public codes, making it accessible for any attacker. A potential attacker tries to find a codeword of $\mathcal{C}'$ of weight $d(\mathcal{C}_\rE)$. The attacker has to find a low-weight codeword (weight $d(\mathcal{C}_\rE) \leq t$) of a code of the same length as the code $\mathcal{C}$. \emph{Information set decoding} (ISD)~\cite{Prange} is a brute forcing decoding algorithm that aims at finding $k$ error-free codeword positions. The decoder then reverts the encoding (encryption) and checks if the obtained word is a valid codeword of $\mathcal{C}$. However, for increasing $t$, the complexity of ISD becomes infeasible for a practical implementation. For the classical McEliece cryptosystem, usually $d(\mathcal{C}_{\rE}) = t$ but for an interleaved McEliece cryptosystem, the attacker can narrow the search for revealing $d(\mathcal{C}_\rE) < t$ erroneous positions. This threat is slightly mitigated by picking $\mathcal{C}_\rE$ as a code with a good minimum distance as in~\cite{lukaslia}, as explained in the following subsection.

\subsection{Support of the Subcode Attack} \label{subseq:support_subcode_attack}
Since the code $\code_\rE$ is a subcode of $\code'$ and moreover each codeword of $\code_\rE$ has at most weight $t$ and shares the same error 
support, it is possible to reveal the error positions by finding the support of the error code $\code_\rE$ \cite{Tillich}. 
For nonbinary codes, it is possible to circumvent this attack by choosing the error matrix as a generator matrix of a code with  large minimum distance~\cite{lukaslia}. This ensures that there will be no error vector with weight less than $d(\mathcal{C}_\rE)$. 

The support of the error matrix $\bfE$ has cardinality $t$. The transmitter chooses a submatrix $\bfE'$  containing  the nonzero columns of the matrix $\bfE$. We define the new code $\code_\rE'$ of length $t$, dimension $\beta$, and minimum distance $d(\code_\rE)$, where $\beta$ has to be optimized. Note that $\beta$ can be much smaller than the interleaving order, thus implying that the error matrix $\bfE$ cannot have full row rank. {The value of $\beta$ has some limitations. If $\beta$ is picked very small and $32$-bit representation of the real numbers is used, an attacker can use brute force  to reveal some error-free combinations of the message. However, for $\beta \geq 4$, the brute force attempt requires to check at least $2^{128}$ possibilities, making it computationally infeasible.}  

Let $\bfG_{\rE}'$ be a generator matrix of the code $\code_\rE'$. For the case where the interleaving order is large, we can modify the error submatrix $\bfE'$ as 
\begin{equation} \label{eq:errorprime}
    \bfE' = 
    \begin{pmatrix}
    \bfG_{\rE}'^\transpose, (\bfS_1 \bfG_{\rE}')^\transpose, \hdots, 
    (\bfS_{\frac{k}{\beta}}\bfG_{\rE}')^\transpose
    \end{pmatrix}^\transpose\,,
\end{equation}
where each of the $\bfS$ matrices is a full-rank square matrix. Due to the linear dependency introduced in $\bfE'$, its row span (the codebook of the code having $\bfE'$ as a generator matrix) is the same as the one from $\bfG_{\rE}'$. A large interleaving order implies some threats from existing decoding algorithms such as in~\cite{MEtznerKapturowski, Combined_block_symbol_error_correction}. However, the work in~\cite{MEtznerKapturowski} is restricted to full-rank error matrices, while the decoding procedure of~\cite{Combined_block_symbol_error_correction} leverages non-full rank error matrices, but the authors do not provide a polynomial-time decoding algorithm for the case where MDPC codes are used.

\subsection{Decoding One-Out-Of Many}\label{subsec:DOOM}
{The decoding one-out-of-many (DOOM) attack~\cite{DOOM} considers the case whereby an illegitimate entity has access to many ciphertext vectors, but it is sufficient to decode only one of them. }
{In~\cite{DOOM}, it is shown that the \emph{work factor} of syndrome decoding is reduced by a factor $\sqrt{\nu}$, where $\nu$ is the number of ciphertexts with linearly independent error vectors available to the illegitimate entity. The reduction of the work factor is only  possible for a given  regime of the parameters. Notably, in~\cite{DOOM}, it is shown that if the number of linearly independent codewords is $\nu \leq \binom{n}{t}$, then the work factor can be reduced. The DOOM attack was originally designed for the binary case and it is not known if the reduction of the work factor holds over larger field sizes. However, we consider it as a \emph{worst-case} scenario and examine this attack in the calculation of the work factor.}

\subsection{Connection to Compressed Sensing} \label{subsec:Discussion}
We consider a code-based cryptosystem over the reals which, to the best  of our knowledge, has not been considered so far. Thus, we need to discuss about special threats that can risk the security of our scheme due to codes over the reals. 
The  problem of decoding codes over the reals has been studied and is known as the \emph{$\ell_0$-norm minimization problem}, 
\begin{align}\label{eq:ell0norm}
    &\quad\min ||\bfe||_0 \\
    &\text{such that } \bfe\bfH = \bfs\,, \nonumber
\end{align}
where $\bfs$ is the syndrome vector computed using a ciphertext $\bfc$ (any row of $\bfC$) and $\bfe$ is any row of the error matrix $\bfE$. The $\ell_0$-norm minimization problem in \eqref{eq:ell0norm} is known to be NP-hard~\cite{NP_hard_l0}, but its relaxation to the $\ell_1$-norm  is solvable in polynomial time. 
\textit{Compressed sensing} studies the conditions that the \textit{measurement matrix} ($\bfH$) should have such that the relaxation to the $\ell_1$-norm provides the solution to~\eqref{eq:ell0norm} and designs such matrices. However, for a given matrix $\bfH$ it is known to be NP-hard  to check if the conditions are fulfilled \cite{Bandeira2013}. 
 \textit{Sparse approximation} also tries to solve the $\ell_0$-norm minimization for a given matrix $\bfH$ by relaxing the $\ell_0$-norm minimization to a $\ell_1$-norm minimization problem. However, most of the results only hold for some strict assumptions such as nonnegative values  \cite{Sparse_google}. All the restrictions that the cited works consider are taken into account while constructing the scheme. 

\subsection{Calculation of the Security Level} \label{subsec:security_level_calc}

We define the security level in bits as $$\text{SL} = \log_2 (\text{WF}_\text{best})\,,$$ 
where $\text{WF}_\text{best}$ is the number of computations needed (work factor) of the fastest existing attack. For this, we will compute the complexity of the fastest version of ISD. 
It is shown in~\cite{Meurer_thesis} that sophisticated ISD algorithms, e.g., Stern~\cite{Stern} and Lee--Brickell~\cite{Lee_Brickell}, perform asymptotically the same as plain ISD  for codes over an alphabet with large cardinality \cite{Prange}. In \cite{peters2010information}, the complexity is computed for codes over $\mathbb{F}_q$, where $q$ is a prime number.

Codes over the reals can be seen as codes over a field with infinite alphabet size. However, an implementation of our scheme would require finite precision. Hence, we assume for the security level computation that our MPDC code operates over $\mathbb{F}_q$, where $q = 2^{32}$ as we consider  $32$-bit precision. 

\section{Numerical Results}\label{sec:num_res}

\ctable[
caption = {The security levels for two choices of $t$},
label = table:seclevels,
width = .5\textwidth,
]{ccccccccc}{
}{
    \FL
    $\log_2(q)$ & $n$ & $k$ & $\beta$ & $t_{\text{tot}}$ & $t$ & $d(\mathcal{C}_E)$ & $M$ & SL \ML \midrule
    \multirow{2}{*}{$32$} & \multirow{2}{*}{$4000$} & \multirow{2}{*}{$2000$} & \multirow{2}{*}{$4$} & \multirow{2}{*}{$256$} & $128$ & $124$ & $16$ & $128$ \NN
    \cmidrule(r){6-9}
    & &  &  &  & $64$ & $60$ & $4$ & $92$ \LL

}

We use an MDPC code of length $n=4000$, dimension $k=2000$, check node degree $90$, and with parity-check matrix constructed using the progressive edge growth algorithm. The public key $\bfG$ requires about $ 30$MB of storage. We pick $t_{\text{tot}} = 256$, for which the value of the frame error rate is $1.33\cdot 10^{-5}$ under peeling decoding. 
We use the constant-weight code construction of~\cite[Ex. 3]{CWC_Vardy} that allows to pick subsets of $\cE$ of cardinality $t$ that  intersect in at most one entry. The number of such available subsets ($M$) and the security level depend heavily on $t$. Since we operate on high field sizes, we consider the complexity of plain ISD~\cite{Meurer_thesis} and we also consider the attack in~\cite{DOOM} as a worst-case scenario (even though this attack is suggested for binary codes only). We tabulate the parameters of our scheme and the underlying security levels in Table~\ref{table:seclevels}.

We test our strategy for a scenario where $N=25$ devices collaborate to train on the MNIST dataset~\cite{mnist}. As proposed in~\cite{intel_journal}, the dataset is preprocessed using kernel embedding via Python's radial basis function sampler of the sklearn library ($2000$ features and kernel parameter $5$) and the labels are one-hot encoded. We assume the preprocessing is performed offline by the devices. The dataset is split into training and test sets. 
 \begin{figure}[t]
     \centering
     \resizebox{!}{.4\textwidth}{\input{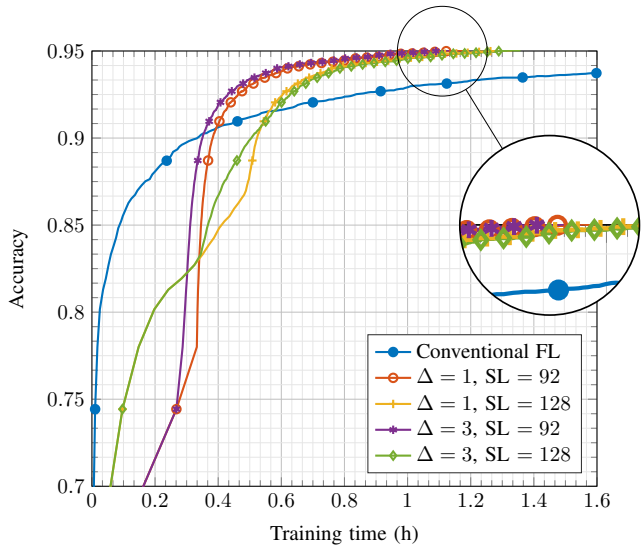}}
     \caption{Training time versus accuracy for the proposed computationally-private  FL scheme and  conventional mini-batch FL. Compared to conventional FL, our scheme yields a significant speed-up for an accuracy level of higher than $95\%$. As expected, a higher security level entails a higher training time.} 
     \label{fig:my_label}
 \end{figure}
We consider learning over single-precision floating point numbers that need $32$ bits. Since the public key $\bfG$ is published in a public database, we assume that every device $i$ computes $\bfA_i = \bfX^{(i)\transpose}\bfX^{(i)}$ and $\bfA_i\bfG$ offline. The learning is started by an all-zero matrix $\boldsymbol{\Theta}^{(1)} = \boldsymbol{0}$ and can thus be computed offline. However, the server asks the devices to send the starting time of their computations to compute the difference in computation time across devices. Thus, we account as the time of the first epoch the time required for the device-to-server communication and the server computation time. Similar to~\cite{schlegel2021codedpaddedfl}, we assume that the random setup time of the devices has a mean of half the deterministic computation time, i.e., $\zeta_i = \frac{\rho_i}{2\tau_i}$. We refer to the system parameters from~\cite{schlegel2021codedpaddedfl}, where $10$ devices have a MAC rate of $25 \cdot 10^6$ MAC/s, $5$ devices have $5 \cdot 10^6$, $5$ other have $2.5\cdot 10^6$, and the last $5$ devices have a MAC rate of $1.25\cdot 10^6$. The central server has a MAC rate of $8.24\cdot 10^{12}$ MAC/s. We refer to the LTE Cat 1 standard for IoT devices where the communication rates are $\gamma^{\text{d}} = 10$ Mbit/s and $\gamma^{\text{u}} = 5$ Mbit/s and the D2D communication link has a bit rate of $\gamma^{\text{D2D}} = 5$ Mbit/s. 
We provide as a benchmark the  conventional mini-batch FL where the batch size is a fifth of the devices' data. We allow the first gradient of the conventional FL to be computed offline for a fair comparison. {The regularization parameter is  $\lambda = 9\cdot10^{-6}$ and the initial learning rate is $\mu = 6$. At the $200$-th and $350$-th epoch we update the learning rate as $\mu \leftarrow \mu \cdot 0.8$.} For every communication, we assume a packet overhead of $10\%$, and a dropout occurs with probability $p = 0.1$. In Fig.~\ref{fig:my_label}, we plot the simulated accuracy versus training time. Compared to conventional mini-batch FL, our best scheme achieves a speed-up  of $4.7$ and $4$  for a security level of $92$ and $128$ bits.  The scheme in~\cite{schlegel2021codedpaddedfl} needs more than $2$ hours just to transfer the data, thus making it impractical for a scenario where D2D links are needed. We also omit the comparison to the scheme in~\cite{intel_journal}, since the privacy of the devices' data is not preserved. Our simulations show that the optimum value of $\Delta$ is $3$ and $1$ for the security level of $92$ and $128$ bits, respectively. Note that the security level of $92$ bits allows $M=16$ distinct error supports, thus allowing to ignore $16$ rounds in the earlier epochs. For the security level of $128$ bits, $M=4$, thus delaying the epochs where the slower devices are ignored. 

The performance of the proposed scheme can be improved by employing a better decoding strategy, such as inactivation decoding~\cite{Inactivation_Fran}.

\section*{Acknowledgment}

The authors would like to thank Reent Schlegel for his helpful comments and providing his simulation code. 

\balance
\bibliographystyle{IEEEtran}
\bibliography{CSFL}

\end{document}